\documentclass[twocolumn]{IEEEtran}          
\usepackage{graphicx}
\usepackage{subfig}
\usepackage{listings}
\usepackage{amsmath}
\usepackage{url}
\begin{document}

\title{City on the Sky: Flexible, Secure Data Sharing on the Cloud}


\author{Dinh Tien Tuan Anh, Wang Wenqiang, Anwitaman Datta
\\
\{ttadinh,wqwang,anwitaman\}@ntu.edu.sg}




\maketitle

\begin{abstract}
  Sharing data from various sources and of diverse kinds, and fusing
  them together for sophisticated analytics and mash-up applications
  are emerging trends, and are prerequisites for grand visions such as
  that of cyber-physical systems enabled smart cities. Cloud
  infrastructure can enable such data sharing both because it can
  scale easily to an arbitrary volume of data and computation needs on
  demand, as well as because of natural collocation of diverse such
  data sets within the infrastructure. However, in order to convince
  data owners that their data are well protected while being shared
  among cloud users, the cloud platform needs to provide flexible
  mechanisms for the users to express the constraints (access rules)
  subject to which the data should be shared, and likewise, enforce
  them effectively. We study a comprehensive set of practical
  scenarios where data sharing needs to be enforced by methods such as
  aggregation, windowed frame, value constrains, etc., and observe
  that existing basic access control mechanisms do not provide
  adequate flexibility to enable effective data sharing in a secure
  and controlled manner. In this paper, we thus propose a framework
  for cloud that extends popular XACML model significantly by
  integrating flexible access control decisions and data access in a
  seamless fashion. We have prototyped the framework and deployed it
  on commercial cloud environment for experimental runs to test the
  efficacy of our approach and evaluate the performance of the
  implemented prototype.
\end{abstract}
\paragraph*{Keywords} cloud computing, access
    control, flexible sharing, fine-grained policies, XACML 

\section{Introduction}
\label{sec:introduction}

The emergence of cloud computing in recent years is rapidly changing
the way businesses and government agencies, as well as individuals,
are storing and managing their data as well as workflows. Instead of
developing and maintaining individual data management infrastructures
and data sharing mechanisms, data owners now leverage on the cloud
services to make their data available to users. The fact that data
from multiple sources now reside in one logical place, i.e., the
cloud, makes it much easier than ever before to develop large scale
applications that require data and knowledge from multiple domains and
sources. These applications could include environmental study, city
infrastructure planning, disaster monitoring, and many more. In an era
when the cloud infrastructure was non-existent, to develop such
applications, the developer would have to first talk to individual
data owners to specifically provide the data to them, which is likely
to involve tedious administration procedures such as signing documents
regarding the privileges and responsibilities of each parties, apart
from the cumbersome process of actually shipping the data. Then the
developer would have to develop software that work with the individual
data exchange interfaces/protocols provided by different owners to
collect and reformat the data before they could be fed into the
applications for analysis or real-time monitoring tasks.

On the multitenant cloud, such data from diverse sources are naturally
collocated, making it much easier and much more efficient for the
application developers to obtain what they need for their work. More
specifically, the storage and data exchange can be handled efficiently
by the cloud providers. This means data owners need not worry about
how to share, but what and who to share.  Putting one's proprietary
data online on the cloud raises concerns regarding data security,
privacy and ownership. Even if the cloud service provider is trusted,
and legally obliged (through service level agreements and law
enforcement) to prevent illegal access of data and information
leakage, there needs to be meaningful, comprehensive and flexible ways
for the data owners to express their sharing preferences, in a manner
which can readily be interpreted and enforced by the cloud service
provider. This paper discusses how this can be achieved. One can
further argue how this can be realized if the cloud service provider
is not even trusted, but that is an issue outside the scope of this
work, and is part of our future work.


The objective of this work is to propose and showcase a framework for
sharing data on the cloud. The framework, called \textbf{eXACML},
facilitates sharing in an \emph{easy-to-use}, \emph{secure},
\emph{flexible} and \emph{scalable} manner. For security, we make use
of/extend XACML~\cite{oasisxacml} --- the XML-based and popular
framework for access control. XACML has become a standard for
specifying and enforcing access control policies. It evaluates
requests for resources against a set of policies and returns
\emph{permit} or \emph{deny} decision, which does not involve
accessing any data. In eXACML, we extend XACML to support more
fine-grained policies as well as to handle data processing. We
demonstrate eXACML's flexibility by using it in different access
control scenarios with different levels of granularity. For usability,
eXACML provides an intuitive, easy-to-use interface for data
owners and data users to specify and enforce security policies and to
access shared data. Finally, we carry out experiments to evaluate the
framework performance in a cloud-like environment, the results of
which suggests that eXACML is scalable. We motivate our work with
scenarios from ongoing works on better city planning, specifically
related to weather and traffic information, and the evaluations are
also based on datasets, part of which are real, while the rest is
synthetic.

In summary, the main contributions of this work are as follows:

\begin{enumerate}
\item We demonstrate the needs for secure and flexible data sharing
  with practical examples involving city planning and management based
  on data from weather and traffic monitoring stations. We discuss
  scenarios in which access control with different levels of
  granularity of data access are needed.

\item We extend the XACML framework to support fine-grained policies.
  In particular, fine-grain access control policies (which require
  data filtering) are expressed within \emph{obligations} that are
  passed from the Policy Decision Point (PDP) to the Policy
  Enforcement Point (PEP), which connects to the database and
  processes the data queries embedded in the obligations. We refer to
  this implementation as\\ \textbf{XACML*}. We discuss why this approach
  could perform better than the traditional approach based on views.

\item We implement a prototype of the framework\\ (\textbf{eXACML}),
  providing additionally, an easy-to-use user interface. The prototype
  allows data owners to easily add and modify their policies. Data
  users can query meta data and details of access policies at remote
  servers. They can also specify aggregated data from multiple sources
  in single requests. Responses to data requests contain information
  of matching policies, enabling flexible conflict resolutions.

\item We evaluate the performance of our prototype in cloud-like
  settings. Our experiments illustrate that the framework incurs
low overhead. We attribute this scalability to the framework's
  ability to cache responses and perform aggregation of responses from
  multiple sources prior to returning them to the data users.
\end{enumerate}

The rest of this paper is organized as follows: Section 2 describes
practical scenarios that motivates our framework. Section 3 details
our extensions to XACML, followed by the logical design of our
framework in Section 4. The prototype and its evaluation are presented
in Section 5. We discuss related and future works in Section 6 and
Section 7 and conclude in Section 8.

Before proceeding further, we'll like to make a final note on the scope of the current work and implementation. Broadly speaking, there are two kinds of data - data already stored in the system (which we refer to as archived/archival data), and data stream, where live data is flowing into the system. Likewise, the queries could be `on demand', typically on the stored data, or continuous queries, to be evaluated on the incoming data streams. The current implementation deals with on demand queries on stored data. This is summarized in Table \ref{tab:scope}.

\begin{table*}
\centering
\begin{tabular}{|c|c|c|}
\hline
Query/Database & Archival (relational) databases & Stream databases\\ \hline
On demand query & current implementation & n/a \\ \hline
Continuous query & n/a & Future work  \\ \hline
\end{tabular}
\caption{Scope of eXACML, regarding database and query type}
\label{tab:scope}
\end{table*}

\section{Motivating Example}
\label{sec:example}

As increasing portion of the world population is rapidly moving to the
cities, while the resources at our disposal are shrinking at an
alarming rate, numerous research and industrial initiatives (e.g.,
IBM's smart cities
initiative \footnote{\url{http://www.ibm.com/smarterplanet/us/en/smarter_cities/overview/index.html}})
are focusing in realizing what are being termed as `smart(er) cities'
in order to manage resources efficiently at the city scale. Enabling
such a move towards smarter cities are cyber-physical systems
aggregating data and actuating the necessary resource management
actions at the edge, while the necessary data storage and analytics is
carried out on cloud based back-end.

In this section, we use some scenarios of road congestion analysis to
showcase the need among data owners for flexible data sharing.

\subsection{Settings.}
Noticing that one of the major
expressways in the city suffers serious congestion during every monsoon
season, Singapore's Land Transport Authority (LTA) has, after preliminary studies, hypothesized that such congestion
is mainly caused by three factors, (1) large number of vehicles on the
road, (2) slow speed of vehicles, (3) bad weather.

To validate such preliminary conclusions and build a traffic condition model during the monsoon
season, researchers need more data. Fortunately, many organizations have been collecting related data: LTA
itself has a number of sensors deployed along the road side to record
traffic volume, i.e., the number of vehicles passing by at unit time; furthermore, another independent entity, a
large local taxi company, collects the speed and location data from
their taxis' GPS devices. At almost any time, there are a number of
such taxis running over the whole stretch of the express way. Likewise, the national environmental agency (NEA)
has several weather stations deployed close to the congested areas,
that record weather parameters such as temperature, humidity, rain
rate, etc.

If all these different data owners use a shared cloud infrastructure\footnote{Note that we are unaware of the current practice of the individual organizations mentioned above, and what follows is a hypothetical scenario.} to store and process the above mentioned data-sets for their individual needs, then when complex analytics involving multiple such datasets become necessary, the data is readily available on the infrastructure thanks to such collocation on the multi-tenant cloud.

Suppose the data are stored in relational tables as shown in Table
\ref{tab:traffic_volume_data} for traffic volume information, Table
\ref{tab:vehicle_speed_data} for cab's location and speed information
and Table \ref{tab:weather_data} for weather information.

\begin{table*}
\centering
\begin{tabular}{l c}
  SamplingTime  & TrafficVolume  \\
  \hline
  2011-06-06 10:00:00 & 60 \\
  2011-06-06 10:05:00 & 67 \\
  2011-06-06 10:10:00 & 50 \\
  ... & ... \\

\end{tabular}
\caption{Table TrafficInfo: Traffic volume data from road side sensors}
\label{tab:traffic_volume_data}
\end{table*}

\begin{table*}
\centering
\begin{tabular}{l c c c}
  SamplingTime  & Speed (km/hr) & latitude & longitude \\
  \hline
  2011-06-06 10:00:00 & 100 & x1 & y1\\
  2011-06-06 10:05:00 & 80  & x2 & y2\\
  2011-06-06 10:10:00 & 40  & x3 & y3\\
  ... & ... \\
\end{tabular}
\caption{Table VehicleInfo: Vehicle speed and location data from GPS devices}
\label{tab:vehicle_speed_data}
\end{table*}

\begin{table*}
\centering
\begin{tabular}{c *{4}{c}}
SamplingTime             & Temperature(C) & Humidity (\%)  & RainRate (mm/hr) &  ... \\
\hline
2011-06-06 10:00:00 & 27.2 & 70 & 0.0  & ...  \\
2011-06-06 10:01:00 & 27.5 & 70 & 0.0  & ...  \\
2011-06-06 10:02:00 & 27.5 & 73 & 0.0  & ...  \\
2011-06-06 10:03:00 & 27.4 & 72 & 0.0  & ...  \\
2011-06-06 10:04:00 & 27.3 & 75 & 0.0  & ...  \\
2011-06-06 10:05:00 & 27.3 & 76 & 0.0  & ...  \\
2011-06-06 10:06:00 & 27.0 & 77 & 0.1  & ...  \\
2011-06-06 10:07:00 & 27.1 & 80 & 5.0  & ...  \\
2011-06-06 10:08:00 & 26.8 & 81 & 14.0  & ...  \\
2011-06-06 10:09:00 & 26.6 & 82 & 20.0  & ...  \\
2011-06-06 10:10:00 & 26.5 & 85 & 34.4 & ...  \\
... & ... & ... & ... & ...  \\

\end{tabular}
\caption{Table WeatherInfo: Weather data from weather stations}
\label{tab:weather_data}
\end{table*}


\subsection{Example 1}
\label{exp:case1}
Suppose that NEA decides to share (possibly for a price) only the rain
rate data with LTA researchers, since other weather parameters such as
temperature and humidity are not expected to affect traffic condition
as much as rainfall does in the context of Singapore, and hence LTA
does not want pay for the temperature or humidity information.
Furthermore, even if the original collected data available with NEA is
for one minute interval, it may want to expose only the data
corresponding to five minute averages to LTA. It may also expose the
more detailed data to its own employees or to other customers.

The first constraint corresponds to the projection operation in the
relational database model and a sample SQL query will be something
like "\emph{select RainRate from WeatherInfo}". The second constraint
can be considered as a sliding window query over a data stream, i.e.,
the time series rain rate data. Standard SQL does not support these
kind of queries well, hence additional operations need to be implemented
on top of the RDBMS query engine. To specify a sliding window query on a
time series data sequence in our scenario, five parameters are needed,
namely, the \emph{starting time}, \emph{ending time}, \emph{window
size}, \emph{window advance step} and \emph{aggregation
function}. The \emph{starting time} and \emph{ending time} are the general
temporal constraints that specify the segment of the data stream to
be returned. The \emph{window size} and \emph{window advance step} decide
the length of the query window and how fast the window is moving along
the data stream. The \emph{aggregation function} includes numerical
functions such as \emph{average()}, \emph{max()}, \emph{min()},
\emph{count()}, etc., which are applied to the data records to
summarize the portion of the data stream within the window.

\subsection{Example 2}
\label{exp:case2}
Consider that the taxi company agrees to help the researchers by
providing their taxis' location and speed data, but the company only
wants to share such information for taxis within some specific regions
in the vicinity of the congested areas being studied, instead of
exposing the information about its whole fleet, which it deems
important business secret not to be exposed to third parties. To
enforce such a constraint, a selection operator is applied to the
longitude and latitude columns to filter out those records that are
not supposed to be shared with the researchers. For the sake of
simplicity, assume that this range is specified by a rectangle with
the geographical coordinate of the upper left vertex as
(a$_{1}$,b$_{1}$) and of the lower right vertex as (a$_{2}$,b$_{2}$),
we can have the corresponding SQL query: \emph{select SamplingTime,
  Speed from VehicleInfo v where v.longitude $=>$ a$_{1}$ and
  v.longitude $<=$ a$_{2}$ and v.latitude $>=$ b$_{2}$ and v.latitude
  $<=$ b$_{1}$}.

\vspace{0.7cm}
To enable the above access contraints in XACML, we make use of the
\emph{obligation} element in \emph{policy} element to specify the
constraints. Fig.~\ref{fig:obligation_1} and
Fig.~\ref{fig:obligation_2} present two examples of XACML obligations
that embed these constraints. In Figure \ref{fig:obligation_1}, line 2
indicates that the permission to perform the sliding window query if
the decision returned from PDP is `permit'. Line 3 indicates that the
aggregation function to be used in the sliding window query is average
calculation. Lines 5 to 8 specify that \emph{starting time} is zero
o'clock of June 6th, 2011, \emph{ending time} is zero o'clock of June
7th, 2011, \emph{window size} is 5 minutes and \emph{window advance
  step} is also of 5 minutes. Line 9 indicates that the sliding window
is applied on SamplingTime column as well, besides on the actual rain
rate data column, which is not shown here within the obligation part.
Line 3 in Figure \ref{fig:obligation_2} shows the selection predicate
to be included in the SQL query to be evaluated on the data table,
which only allows vehicle information to be returned if the vehicle's
location is within a given boundary.

\subsection{Fine-grained Policies}
The examples above demonstrate real needs for an access control model
that supports fine-grained policies involving fine-grained data
processing.  At a high level, the models need to be able to express
and enforce the following types of policies:

\begin{enumerate}
\item Aggregated data: Only results of aggregation
  functions over raw data such as \emph{average},\emph{sum},
  \emph{min}, \emph{max} are shared.

\item Trigger-based: a row of data is accessible
  only if the value of a column satisfies a certain predicate:
  exceeds a specific threshold, or is contained within a range. As an
  example, a taxi company is granted access to temperature reading
  only if the temperature is over $30^o$C.

\item Sliding window: a sliding window is specified
  by its starting time, ending time, window size and advance
  step. Only aggregated data (average, for instance) over the windows
  are accessible.

\item Approximation: only data whose values
  approximate those given in the requests are accessible. For example,
  a request includes a value $X$, and the policies is specified such
  that a row of data is returned only if the column $c$'s value $V$
  satisfies $|V-X| < \epsilon$ for some distance function.
\end{enumerate}

We next explore how such fine-grained policies can be flexibly supported.

\section{Flexible Sharing Through Fine-Grained Policies}
\label{sec:flexiblePolicies}

Existing frameworks, such as XACML, do not natively support different levels of granularity to support fine-grained access control.  Nevertheless, XACML has emerged in recent years as a mature and widely used
model for expressing and enforcing access control
policies.  Therefore, we extend XACML in order to support
fine-grained policies, including those described in Section 2.

For the rest of this paper, we assume relational databases (SQL
types) are used for managing data in the back-end. Without loss of generality, but for the purpose of simplicity of exposition, we consider that each database consists of a single table indexed by time values. When requesting for
data, the user provides his credentials (for example, name and role) and specifies the location of data. The response contains either a
\emph{deny} decision (i.e. no access to the data), or permit decision together with the returned data as specified in the policies.

\subsection{XACML}
XACML is an OASIS framework for specifying and enforcing access
control~\cite{oasisxacml}. It is XML based and the latest version
is $3.0$. XACML allows administrators to control their resources by
writing policy files, which are then loaded into a Policy Decision
Point (PDP) module.  An user wishing to access a specific resource
sends request to a Policy Enforcement Point (PEP) where the decision
is made by consulting the PDP. XACML specifies standards for writing
policies, requests and interpreting the response.

\begin{enumerate}
\item \emph{Subjects}, \emph{Resources} and \emph{Actions}. A
  \emph{subject} in XACML has a set of credentials such as its name,
  role, etc. The subject wishes to perform certain \emph{actions}
  (read, write, for example) on a set of system \emph{resources}.
	
\item \emph{Requests}. Request for accessing system resources are
  written in XML. The subject credentials, system resources and
  actions are specified in one or more \emph{Attribute} elements
  included in the Subject, Resource and Action elements respectively.
  Fig.~\ref{fig:requestExample} shows an example of an XACML request
  from a subject with role \emph{admin} to perform \emph{read} action
  the \emph{temperature} column from \emph{weather\_data} database.
	\begin{figure*}
	\scriptsize
        \centering
	\begin{verbatim}
	<Subject>
	   <Attribute AttributeId = ``exacml:subject:role-id''
                      DataType={http://wwww.w3.org/2001/XMLSchema#string}>
	      <AttributeValue>admin</AttributeValue>
	   </Attribute>
	</Subject>
	
	<Resource>
	   <Attribute AttributeId = ``exacml:rdmb-database-id''
                      DataType={http://www.w3.org/2001/XMLSchema#string}>
              <AttributeValue>weather_data</AttributeValue>
           </Attribute>
   	   <Attribute AttributeId = ``exacml:rdmb-column-id''
                      DataType={http://www.w3.org/2001/XMLSchema#string}>
              <AttributeValue>temperature</AttributeValue>
           </Attribute>	
	</Resource>

	<Action>
	   <Attribute AttributeId = ``exacml:action-id''
                      DataType={http://www.w3.org/2001/XMLSchema#string}>
              <AttributeValue>read</AttributeValue>
           </Attribute>
	</Action>	
	\end{verbatim}
	\caption{Example of a well-formed XACML request, in which the
          user with the role \emph{admin} requests \emph{read} access
          to the column \emph{temperature} of the database
          \emph{weather\_data}}
	\label{fig:requestExample}
	\end{figure*}
	
      \item \emph{Policies}. A policy contains a \emph{Target}, a set
        of \emph{Rules} each of which has at most one
        \emph{Condition}, and a set of \emph{Obligations}. Multiple
        policies can be grouped into a \emph{policy set}, which has
        its own Target element. The policy is indexed by its Target
        element, which consists of a number of conditions needed to be
        satisfied by the request before the rest of the policy can be
        evaluated. Conditions are essentially boolean expressions over
        the values included in the request. The policy returns access
        control decision which is either \emph{Permit}, \emph{Deny},
        \emph{Not Applicable} or \emph{Intermediate}. The last two are
        used when there is no applicable policy or an error occurred
        during evaluation. Fig.~\ref{fig:xacmlPolicy} illustrates an
        example of an XACML policy that grants access to subjects with
        \emph{government} role to the \emph{samplingtime} and
        \emph{temperature} columns of \emph{weather\_data}.

	\begin{figure*}
          \centering
          \scriptsize
          \begin{verbatim}
          <Target>
            <Subjects>
              <Subject>
                <SubjectMatch MatchId="urn:oasis:names:tc:xacml:1.0:function:string-equal">
                  <AttributeValue DataType="http://www.w3.org/2001/XMLSchema#string">
                    government
                  </AttributeValue>
                  <SubjectAttributeDesignator AttributeId="exacml:subject:role-id"
                                       DataType="http://www.w3.org/2001/XMLSchema#string"/>
                </SubjectMatch>
              </Subject>
           </Subjects>
           <Resources>
              <Resource>
                <ResourceMatch MatchId="urn:oasis:names:tc:xacml:1.0:function:string-equal">
                  <AttributeValue DataType="http://www.w3.org/2001/XMLSchema#string">
                    weather_data
                  </AttributeValue>
                  <ResourceAttributeDesignator AttributeId="exacml:rdbms-database-id"
                                       DataType="http://www.w3.org/2001/XMLSchema#string"/>
                </SubjectMatch>
             </Resource>
           <Resources>
           <Actions>
             <AnyAction/>
           </Actions>
          </Target>
          <Rule RuleId="example" Effect="Permit">
            <Target> <Subjects> <AnySubject/> </Subjects>
                   <Resources> <AnyResource/> </Resources>
                   <Actions> <AnyAction/> </Actions>
            </Target>              		
            <Condition FunctionId="urn:oasis:names:tc:xacml:1.0:function:string-subset">
              <ResourceAttributeDesignator AttributeId="exacml:rdbms-column-id"
                                      DataType="http://www.w3.org/2001/XMLSchema#string"/>
              <Apply FunctionId="urn:oasis:names:tc:xacml:1.0:function:string-bag">
                  <AttributeValue DataType="http://www.w3.org/2001/XMLSchema#string">
                                  samplingtime
                  </AttributeValue>
                  <AttributeValue DataType="http://www.w3.org/2001/XMLSchema#string">
                                  temperature
                  </AttributeValue>
              </Apply>
            </Condition>				
          </Rule>
          \end{verbatim}

          \caption{Example of a well-formed XACML policy which grant
            access to column \emph{samplingtime} or \emph{temperature}
            of the database \emph{weather\_data} to any subject with
            role \emph{goverment}}

          \label{fig:xacmlPolicy}
        \end{figure*}
	
	When more than one rules are applicable to a particular
        request, they are evaluated according to \emph{rule
          combination algorithm} specified in the policy. Similarly,
        multiple applicable policies in a policy set are evaluated
        according to a specified \emph{policy combination
          algorithm}. Examples of combining algorithms (for both
        policies and rules) are \emph{Permit-overrides} where a permit
        policy or rule is evaluated, and \emph{First-applicable} where
        the first applicable policy is evaluated.
		
      \item \emph{Policy Enforcement Point (PEP)}. User requests first
        go through the PEP, which translates them into canonical forms
        before passing to the PDP.  Additionally, PEP also interprets
        responses and obligations returned from the PDP.  In summary,
        PEP deals with application logics and acts as the access
        control enforcement mechanism. Our framework extends PEP to
        provide support for more fine-grained policies.

      \item \emph{Policy Decision Point (PDP)}. Data owners' policies
        are \emph{loaded} into the PDP, which evaluates requests
        received from the PEP against the active policies. Its main
        task is to efficiently find applicable policies for a given
        request and to quickly evaluate their rules and conditions to
        determine the access control decision. It sends back to PEP a
        well-formed response containing a decision and a set of
        obligations.
\end{enumerate}

\subsection{View-Based vs Obligation-Based}
The traditional access control model in relational databases is based
on \emph{view} \cite{ramakrishnan02}. Basically, a view is the result
of a SQL query on existing tables, to which read/write access are
specified. The database management systems maintain the views and
enforce access control rules on them.

A simple approach based on view to support fine-grained policies with
XACML can be realized as follows. First, views are created with no
access control restriction, and assigned with unique resource IDs.
This can handle all types of policies discussed earlier. PEP maintains
a mapping between the IDs and actual views. Next, the IDs are used to
specify the resources in XACML policies, as well as to construct data
requests.  Once PDP returns a permit decision, PEP retrieves and
returns the corresponding views.

However, there are a number of weaknesses with this approach:
\begin{itemize}
\item Views need to be created prior to policies or requests.
  They must also be removed explicitly by the data owner.

\item Views are static and may be very large in number (potentially
  infinite number of views for trigger-based and sliding window policies).
  Maintaining these views are inefficient at best and impossible at
  worst.

\item An user requesting for data must also maintain a mapping of all
  the view IDs they wish to access. Not only is such a requirement undesirable for data users, but also it is expensive to implement.
\end{itemize}

\begin{figure}
\centering
\includegraphics[scale=0.8]{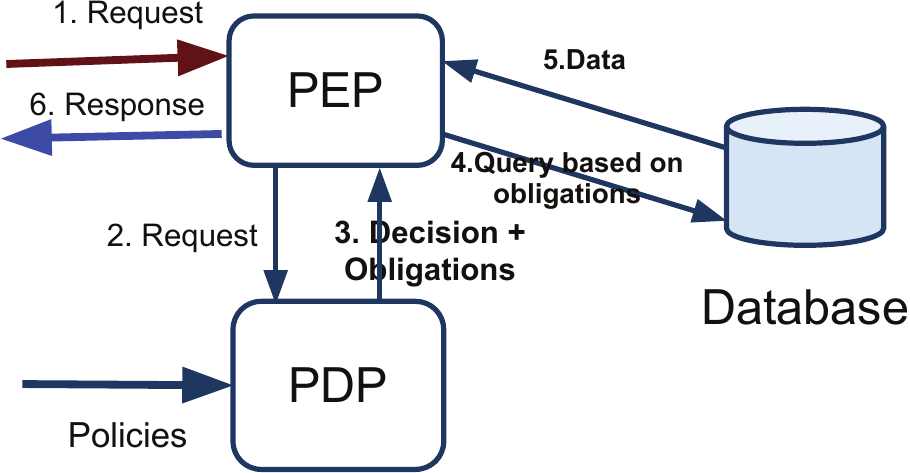}
\caption{Extensions to XACML that support more flexible access control
  policies.}
\label{fig:xacmlExtension}
\end{figure}

Fig.~\ref{fig:xacmlExtension} illustrates the obligation-based
approach (extensions to XACML is highlighted in bold). The basic idea
is to embed queries for creating views into obligations. The PEP, upon
receipt of the obligations, executes the embedded queries on the
database and returns the results in a well-formed response. Unlike the
view-based approach, the size of data (views) maintained by PEP is
bounded. Furthermore, popular queries can be cached by the database management
system or the PEP. In the experiment section, we demonstrate the
benefit of caching in improving request time.

\subsection{Implementations}
\subsubsection{Obligations.}
\begin{table}
\begin{tabular}{l|l}
\hline
\textbf{Description} & \textbf{ObligationId} \\
\hline\hline
Column aggregation & \texttt{exacml:obligation:column-aggregation} \\
Simple selection & \texttt{exacml:obligation:simple-selection} \\
Sliding window & \texttt{exacml:obligation:column-sliding-window} \\
Approximation & \texttt{exacml:obligation:column-approximation} \\
\hline
\end{tabular}
\caption{Obligation types}
\label{tab:obligations}
\end{table}
Using obligation-based approach, policy writers utilize different
types of obligations to specify different database queries. Our
current implementation supports four types of obligations
(Table~\ref{tab:obligations}):
\begin{enumerate}
\item \emph{Column aggregation}: consists of a string attribute with ID \\
  {\small\texttt{exacml:obligation:aggregation-id}}. The string represents an
  aggregation function, such as average (Fig.~\ref{fig:obligation_1},
  line 2-3), min, max, count or sum.

\item \emph{Simple selection}: consists of a string attribute with ID  \\
  {\small\texttt{exacml:obligation:selection-id}}. The string is a boolean
  expression that will be used as the WHERE clause when constructing
  the database query. An example of this obligation is shown in
  Fig.~\ref{fig:obligation_2}, in which the policy restricts access to
  data to within a certain geographical region.

\item \emph{Sliding window}: we assume that the column from which the sliding
  windows are based is of type DateTime (although sliding windows
  could be constructed from any other sortable types). The obligation
  consists of a number of attributes:
\begin{itemize}
\item Sliding window column: string attribute with ID \\
  {\small\texttt{exacml:obligation:sliding-window-column-id}} specifies the
  column of type DateTime from which sliding windows are constructed.

\item Start and End: time attributes with IDs \\
  {\small\texttt{exacml:obligation:sliding-window-start-id}} and \\
  {\small\texttt{exacml:obligation:sliding-window-end-id}} respectively.
	
\item Window size: integer attribute with ID  \\
  {\small\texttt{exacml:obligation:sliding-window-size-id}} specifies the
  window size (in hours).

\item Advance step: integer attribute with ID \\
  {\small\texttt{exacml:obligation:sliding-window-step-id}} specifies how the
  sliding window advances, i.e. the number hours between starting time
  of two consecutive windows.
\end{itemize}
Fig.~\ref{fig:obligation_1} (line 4-10) shows an example of a sliding
window based on \emph{SamplingTime} column. The window's size is 5
hours, starting from \texttt{2011-06-06 00:00:00}, advancing in 5-hour
steps until \texttt{2011-06-07 00:00:00}.

\item \emph{Approximation}: this obligation specifies the acceptable distance
  between the column values with respect to the values included in the request.
  Attributes containing column IDs are specified in both the requests
  and the policies. Specifically:
\begin{itemize}
\item In the request: string attribute with ID\\
  {\small\texttt{exacml:data-value-id}} is of the form\\
  {\small\texttt{<columnId>:<value>}} which represent the value of the
  specified column.

\item In the policy: string attribute with ID \\
  {\small\texttt{exacml:obligation:approximation-param-id}} contains the
  column IDs. Columns specified in the requests must be a subset of
  what is specified in the policies. Also
  required is a double attribute with ID \\
  {\small\texttt{exacml:obligation:approximation-value-id}} which represents
  the distance between the vector of column values in the database and
  that included in the request.
\end{itemize}

\end{enumerate}

\begin{figure*}
\centering
\scriptsize
\begin{verbatim}

<Obligations>
  <Obligation ObligationId="exacml:obligation:column-aggregation" FulfillOn = "Permit">
    <AttributeAssignment AttributeId="exacml:obligation:aggregation-id"
                         DataType = "http://www.w3.org/2001/XMLSchema#string">
       avg
    </AttributeAssignment>
  </Obligation>
  <Obligation ObligationId="exacml:obligation:column-sliding-window" FulfillOn = "Permit">
    <AttributeAssignment AttributeId="exacml:obligation:sliding-window-start-id"
                         DataType = "http://www.w3.org/2001/XMLSchema#time">
       2011-06-06 00:00:00
    </AttributeAssignment>
    <AttributeAssignment AttributeId="exacml:obligation:sliding-window-end-id"
                         DataType = "http://www.w3.org/2001/XMLSchema#time">
       2011-06-07 00:00:00
    </AttributeAssignment>
    <AttributeAssignment AttributeId="exacml:obligation:sliding-window-size-id"
                         DataType = "http://www.w3.org/2001/XMLSchema#integer">
       5
    </AttributeAssignment>
    <AttributeAssignment AttributeId="exacml:obligation:sliding-window-step-id"
                         DataType = "http://www.w3.org/2001/XMLSchema#integer">
       5
    </AttributeAssignment>
    <AttributeAssignment AttributeId="exacml:obligation:sliding-window-column-id"
                         DataType = "http://www.w3.org/2001/XMLSchema#string">
       samplingtime
     </AttributeAssignment>
  </Obligation>
</Obligations>
\end{verbatim}
\scriptsize
\caption{Obligation portion of the XACML policy for Example \ref{exp:case1}}
\label{fig:obligation_1}

\end{figure*}

\begin{figure*}
\centering
\scriptsize
\begin{verbatim}
<Obligations>
  <Obligation ObligationId="exacml:obligation:simple-selection" FulfillOn = "Permit">
    <AttributeAssignment AttributeId="exacml:obligation:selection-id"
                         DataType = "http://www.w3.org/2001/XMLSchema#string">
       longitude >= a1 and longitude <= a2 and latitude >= b2 and latitude <= b1
    </AttributeAssignment>
  </Obligation>
</Obligations>
\end{verbatim}
\caption{Obligation portion of the XACML policy for Example \ref{exp:case2}}
\label{fig:obligation_2}
\end{figure*}

\subsubsection{Handling obligations.}
PEP extracts attributes embedded in the obligations and constructs
corresponding queries to be executed on the database. It is not
uncommon for a policy to have more than one types of obligations,
which allows for more expressive, fine-grained conditions for
accessing data. Essentially, PEP creates queries of the following
form:
{\small
\begin{align}
  \mbox{\texttt{select }} & \mbox{\texttt{f(column\_1),
      f(column\_2),..,f(column\_n)}} \notag\\ 
      & \mbox{\texttt{from Table\_name where Where\_Condition}}
\label{eq:typicalDatabaseQuery}
\end{align}
}
where \texttt{column\_i} $(1 \leq i \leq n)$ and \texttt{Table\_name}
are extracted from the Resources element of the request. When no
obligation is returned, \texttt{f} and \texttt{Where\_Condition} are
set to empty strings. In this case, the query becomes:
{\small
\begin{verbatim}
select column_1, column_2,..,column_n
       from Table_name
\end{verbatim}
}
PEP obtains \texttt{f} from the string attribute in the column
aggregation obligation. When a simple selection obligation is
returned, \texttt{Where\_condition} is taken directly from its string
attribute. For approximation obligations, the PEP first retrieves a
vector of values from the request, namely $(x_1, x_2, .., x_k)$ from
columns\\ $c_1, c_2, .., c_k$. It then obtains the distance value
$\delta$ in the obligation, and sets \texttt{Where\_condition} as:
\[
\mbox{\texttt{sqrt$((c_1-x_1).(c_1-x_1) + .. + (c_k-x_k).(c_k-x_k)) < \delta$}}
\]

Handling sliding-window obligations are more complex. First, the tuple
\\ $(start, end, window\_size, advancing\_step)$ are extracted from
the obligation. The total number of windows are:
\[nW = \lfloor \frac{end-start-window\_size+1}{advancing\_step}\rfloor
+1\] For every window, PEP creates a different query.  More
specifically, let $c$ be the column (of type DateTime) from which the
sliding windows are constructed, a query $i$ $(0 \leq i < nW)$ is of
the form:
{\small
\begin{align*}
  \mbox{\texttt{select }} & \mbox{\texttt{f(column\_1),
      f(column\_2),..,f(column\_n)}}\\
  & \mbox{\texttt{from Table\_name where Where\_Condition}}\\
  & \qquad \mbox{\texttt{AND c $\geq$ start+step*i}}\\
  & \qquad \mbox{\texttt{AND c < start+step*i+size}}
\end{align*}
}
where \texttt{Where\_Condition} are constructed from simple selection
and approximation obligations.


\section{The Logical Framework}
\begin{figure}
\includegraphics[scale=0.55]{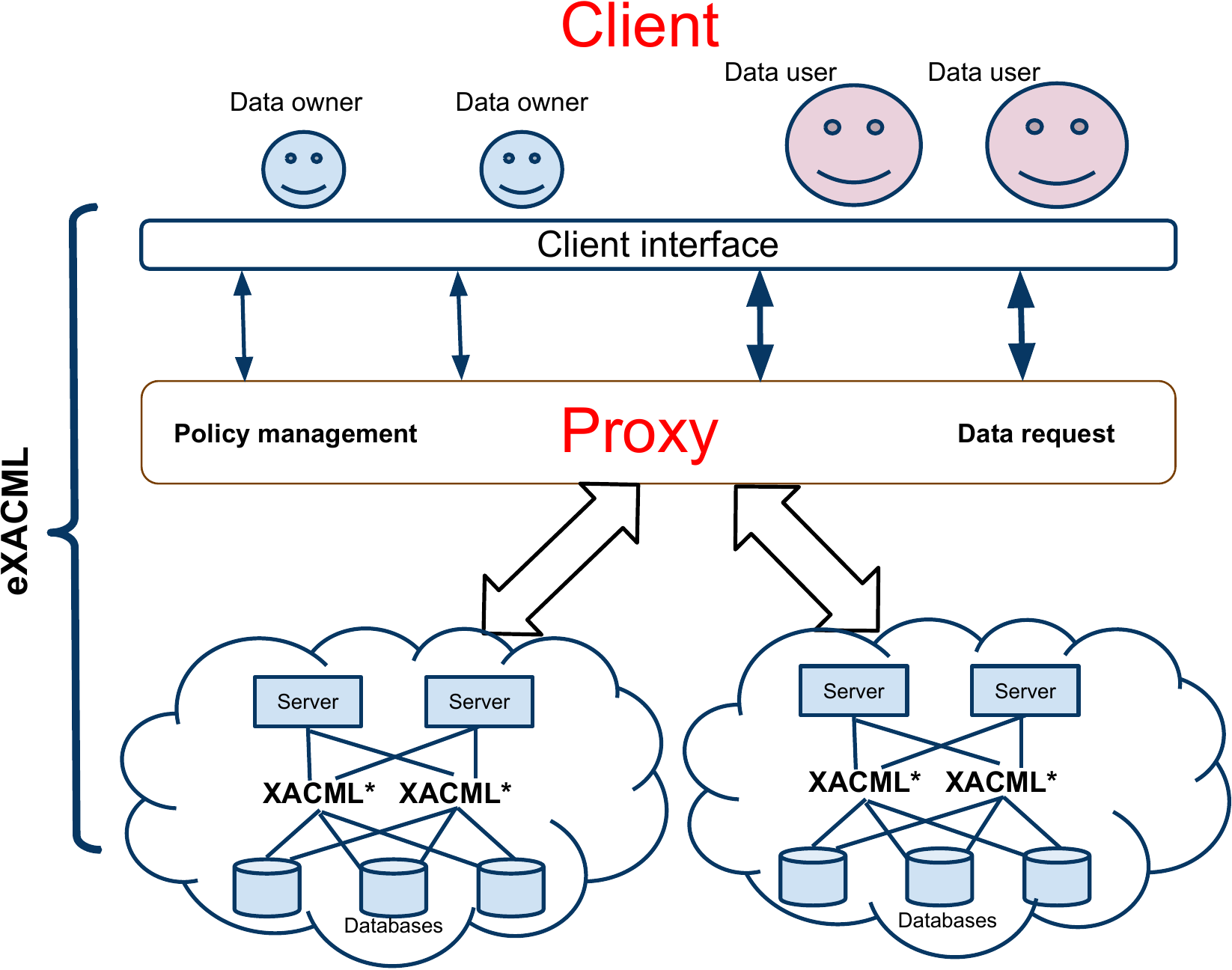}
\caption{eXACML framework. XACML* denotes the extended
 XACML described in Section 3. }
\label{fig:framework}
\end{figure}
This section presents our design of the framework that enables secure,
easy-to-use, flexible and scalable data sharing. The security comes
from the use of XACML for specifying and enforcing access control. The
flexibility property is the result of our enhancement to XACML which
supports a wider range of access control policies. Usability and
scalability are achieved through a simple client interface and the use
of a proxy server, whose details are described below.

\subsection{Entities}
Fig.~\ref{fig:framework} illustrates the main entities and how they
interact in our framework. \emph{Clients} consist of data owners who
wish to share and enforce access control on their datasets, and of
data users who are interested in accessing the data. A data owner can
have more than one datasets and a data user can request access to
multiple datasets.  \emph{Databases} are database servers which manage
clients' datasets. Access to the database is controlled by at least
one instance of XACML* (discussed below). These servers are likely to
be remote and maintained by a third party (cloud) provider.

Our framework --- \textbf{eXACML} --- is positioned in between clients
and databases (Fig.~\ref{fig:framework}). Its roles are to mediate
their interactions and to safeguard the databases.  Essentially,
eXACML is made up of a client interface, a proxy server, cloud
servers and XACML* instances.
\begin{itemize}
\item Clients interact with the databases through a local \emph{client
    interface} that parses inputs into request messages and forwards
  them to the proxy server. It waits and interprets response messages
  before returning them back to the clients. This interface abstracts
  out the complexity of exchanging well-formed messages with the proxy
  server. It allows clients to share and query data in an intuitive
  manner.

\item A \textbf{cloud server} (or \emph{server}), usually located in
  the same machine as the databases, accepts and processes client
  requests. We will refer to this component as \emph{server}. It
  manages and responses to meta queries concerning XACML* instances.
  For data requests, it forwards them to the appropriate XACML*
  instances and sends the results to the proxy in well-formed messages.

\item \textbf{XACML*} is an implementation of the extended XACML model
  described in Section 3 (Fig.~\ref{fig:xacmlExtension}). It processes
  data requests (received from the cloud server) by first asking PDP
  for the access decision. If permitted, it executes the obligations,
  which involves querying the database. The result is forwarded back
  to the cloud server.

\item Communications between clients and servers go through a
  \textbf{proxy server} (or \emph{proxy}). It processes requests from
  clients before forwarding them to the servers, and combines the
  results into client response messages.  As an example, suppose a
  request from a data user requires accessing data from multiple
  datasets, the proxy first creates multiple requests and sends to the
  corresponding servers. It waits for all the responses from servers,
  then combines the results into a single response message for the
  data user.

  The benefit of having the proxy server is two-fold:
	\begin{enumerate}

	\item \emph{Improved performance}: Combining data before
          returning to the users reduces communication costs. Caching
          at the proxy can also improve response time and reduce both
          computation and communication costs for the database
          servers. We demonstrate this effect in the evaluation
          section.

	\item \emph{Additional level of abstraction}: The proxy server
          acts like a DNS service mapping datasets into to global,
          easy-to-remember names, achieving network data independence, which makes it easier for clients to
          manage and query data.
	
	\end{enumerate}

\end{itemize}

\subsection{Trust and Data Model}
We assume cloud severs and the proxy server are \emph{honest}. This means that
they are trusted to run the correct, latest eXACML framework. They
are also trusted not to violate data privacy. More specifically, the
proxy is trusted not to tamper with the data received from database
servers, and not to violate data privacy. The only \emph{adversaries}
are rouge clients who can collude in attempt to gain unauthorized
access to the datasets belonging to honest data owners. We remark that
these assumptions (particularly, that of trusted service providers) are reasonable since cloud service providers are striving to
gain reputation to run their business, and furthermore have legal obligations based on Service Level
Agreements~\cite{popa11}.

We assume that datasets are managed by relational database systems.
For simplicity, each data owner has at most one dataset. This
assumption can be relaxed by \emph{virtualizing} the data owner, so
that it has multiple identities, each of which possesses a different
dataset.

\subsection{Cloud Model}
We now discuss different ways to connect the database, XACML* and
cloud server components. As seen in Fig.~\ref{fig:framework}, the
number of servers, the number of databases and XACML* instances do not
have to match. In particular, multiple databases may share the same
XACML* instance, while a cloud server may handle multiple XACML*
instances.

A server represents a logical, addressable machine to which the proxy
connects. One server can handle requests for multiple datasets, but we
assume each server is connected to one dataset. This assumption is
reasonable since each data owner has at most one dataset, and it is
likely that data owners use independent virtual machines.

Next, we consider the question of how XACML* instances are shared
among databases.  At one extreme, a single XACML* instance is
sufficient to deal with all access requests. In this case, the servers
connect to the the same XACML* instance, and policies are added to the
same PDP. The PEP has access to multiple databases at different
machines.  However, this approach introduces a single point of
failure, and data owners may prefer to have their access control
systems separated from each other. Moreover, extra layers of
authorization is required to prevent rouge clients from uploading
policies associated with datasets of honest data owners. At the other
extreme, the server maintain one XACML* instance per dataset.  Since
data requests can be processed in parallel, this approach could lead
to significant improvement in performance. However, a potential
drawback is the overhead in maintaining a large number of XACML*
instances, especially if many are idle.

When multiple datasets share the same physical machine (but are in
separate virtual machines), it makes more sense for them to share one
XACML* instance. This approach benefits from the parallelism in
processing requests, while having reduced overhead in maintenance.
However, sharing an XACML* instance experience the same problem with
single point of failure and extra layer of authorization as with a
single XACML* instance.

\begin{figure}
\centering
\includegraphics[scale=0.5]{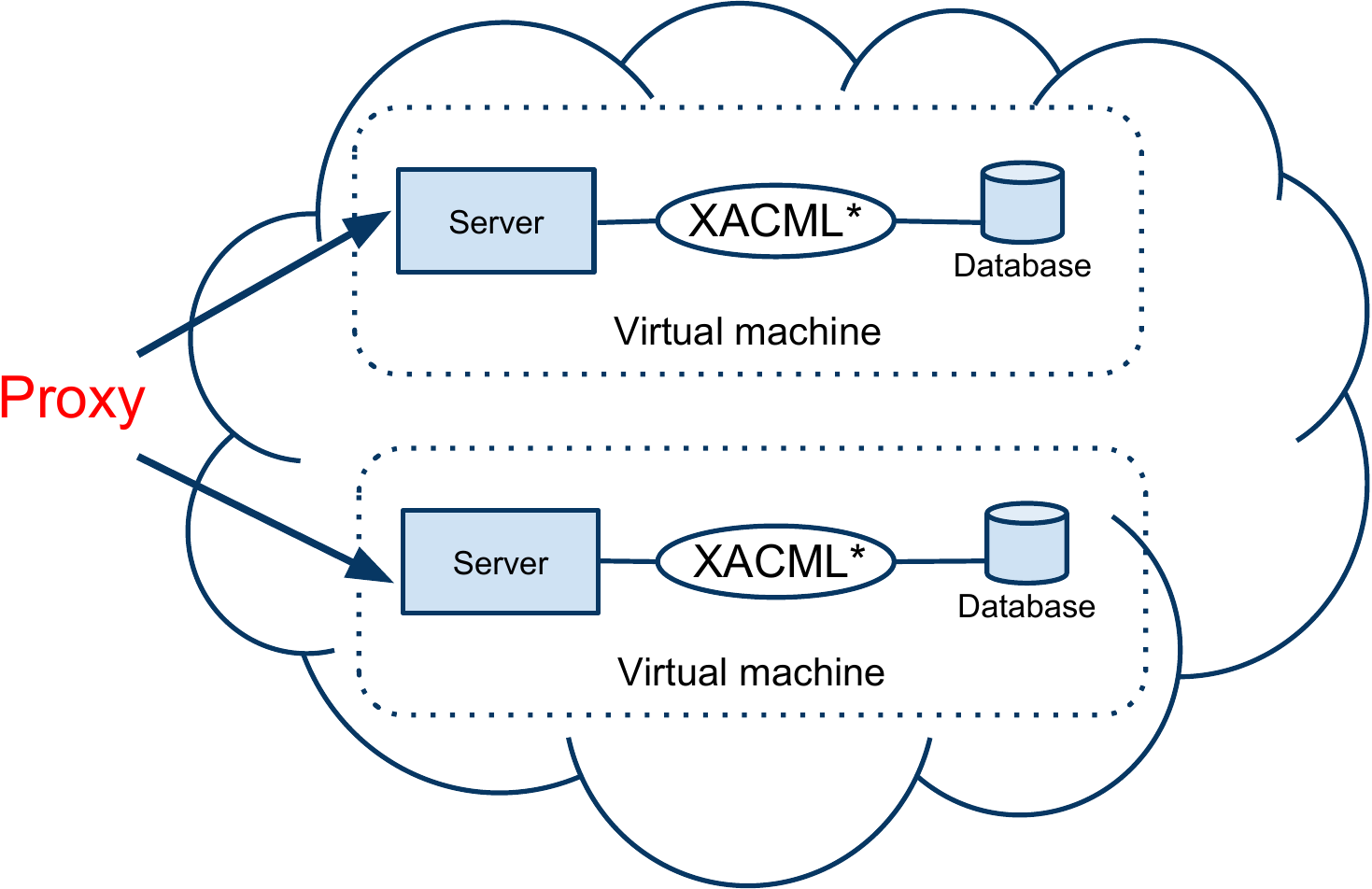}
\caption{Interaction model of the cloud server, XACML* and database }
\label{fig:virtualMachines}
\end{figure}

Considering the above trade-offs, in this paper, we finally adopted the simple, no-sharing approach, i.e. one server
connects to one XACML* that safeguards one database (illustrated in
Fig.~\ref{fig:virtualMachines}). This model does not require another
layer of authorization and therefore is easy to implement.

\subsection{One or Multiple Proxies?}
Having multiple proxies addresses the trust problem associated with a
single proxy. It could also improve client throughputs, since requests
can be processed in parallel. However, joining data --- one of the
proxy's main features --- across multiple proxies is more complex.
Since proxies also maintain data caches, a mechanism for cache
coherence among distributed servers is also required. Therefore,
trade-offs between efficiency and maintenance overhead must be
carefully considered. Our current framework employs only one proxy. We
defer the protocols with multiple proxies for future work.

\subsection{Initialization}
In the beginning, a data owner creates a database for its datasets and
initializes an XACML* instance at a remote data server. The XACML*
instance starts with an \emph{initial policy} specifying who can add
and remove data and policies. This process is done by invoking

{\small
\begin{verbatim}
{success,fail} 
   <- initDatabase(host, port, dataID,
                   databaseType, credentials)
\end{verbatim}
}
where \texttt{host, port} are the address of the server,
\texttt{dataID} is the unique identifier of the dataset,
\texttt{databaseType} is name of the database management system
(MySQL, for example), and \texttt{credentials} consists of the data
owner's name, role and other authentication information for accessing
the server. The client interface wraps these parameters into a message
forwarded to the proxy, then sends it to the specified server.  After
authenticating the data owner, the server creates the database, starts
an XACML* instance and connects its PEP to the database.  Finally, the
server uploads a \emph{root policy} to the newly created XACML*
instance. The root policy specifies that only users with
\texttt{credentials} can add new data, upload new and remove existing
policies. This policy prevents other clients from adding their own
policies to this XACML* instance.

If successful, the proxy creates a new mapping from \texttt{dataID} to
the dataset, as explained next.

\subsection{Data and Policy Management.}
Once a database is initialized successfully, it can be identified
uniquely by its \texttt{dataID}. The proxy maintains a mapping
\texttt{dataID\_to\_desc}, which is a list of:
{\small
\[
\mbox{\texttt{dataID:<host, port, database name>}}
\]
}
All client requests contain \texttt{dataIDs}. The proxy resolves
locations of the dataset using its mapping, before forming new requests
and forwarding them to the appropriate database servers.

\paragraph{Adding and removing data.}
To add or remove new data from a dataset, the data owner invokes
{\small
\begin{verbatim}
{succses, fail} 
    <- addData(data file, dataID, credentials)
{success, fail} 
    <- removeData(remove query, dataID, 
                  credentials)
\end{verbatim}
}
where \texttt{data file} contains data to be added to \texttt{dataID}
using the given \texttt{credentials}. \texttt{remove query} is the
query to remove records from the database. The client interface sends
a request to the proxy, which in turn constructs and forwards a
well-formed XACML request together with the file hash or query hash to
the server. The server keeps the hash as the \emph{pending add} or
\emph{pending removal} token. Only if the access control decision is
`permit' does the client interface sends \texttt{data file} or
\texttt{remove query} to the server, which verifies that the content
hash matches with the \emph{pending add} or \emph{pending remove}
before performing the query.  In this protocol, the hash value is used
to prevent other data owners from adding rouge data or remove
unauthorized data.

\paragraph{Loading and removing policy. }
Every loaded policy is identified uniquely by its ID of the form
\texttt{dataID:policyID} where \texttt{policyID} is the integer index
of the policy. The XACML* instance maintains an index counter which
advances whenever a new policy is added.

To add or remove a policy, a data owner invokes
{\small
\begin{verbatim}
{policyID, fail} 
      <- loadPolicy(policy file, dataID, 
                    credentials)
{success, fail} 
      <- removePolicy(policyID, dataID, 
                    credentials)
\end{verbatim}
}
where \texttt{policy file} contains the XACML file to be uploaded to
\texttt{dataID} using the given \texttt{credentials}. The policy to be
removed is identified by the tuple (\texttt{dataID, policyID}). The
client interface forwards a request to the proxy, which creates a
well-formed XACML request (for loading or removing policy) using
\texttt{dataID} and the \texttt{credential}. Once arrived at the
server, the request is evaluated by the appropriate XACML* instance.
Only if the decision is permit is the \texttt{policy file} added or
the policy \texttt{dataID:policyID} is removed from the corresponding
PDP. In case of policy addition, the new policy ID --- the current
index counter's value --- is forwarded back to the data owner. We
assume that policy is small, thus there is no need for the 2-step
protocols as in adding and removing data.

\paragraph{Querying policy. }
Both data owner and the server keep track of the policy IDs associated
with the dataset. One can query about the loaded policies for a
dataset, using
{\small
\begin{verbatim}
{{(policyID, description)}, fail} 
        <- queryPolicy(dataID, credentials}
\end{verbatim}
}
which returns a set of tuples \texttt{(policyID, description)} where
\texttt{description} is the Description element of the corresponding
policy.

\subsection{Data Request.}
A data user issues a request for data through the client interface.
The request may involve accessing multiple datasets. The data user
knows \texttt{dataIDs}, but may not know of the detailed structure of
the datasets.

\subsubsection{Querying meta data.}
A data user can issue a query for the dataset's meta data prior to
requesting the raw data. Typical meta data includes table names and
schemas. Data owners can restrict access to such information through a
set of policies. To query meta data, the data user invokes:
{\small
\begin{verbatim}
{{tableID}, fail} 
      <- queryTables(dataID, credentials)
{(columnID, type)}, fail} 
      <- queryColumns(dataID, tableID, 
                      credentials)
\end{verbatim}
}
The proxy translates the client request into a well-formed, standard
XACML request in which the \emph{Action} attribute is set to
\texttt{show\_table} or \texttt{show\_column} respectively. If the PDP
returns a permit decision, the PEP retrieves and returns the
database's metadata accordingly.  The result for \texttt{queryTables}
(if permitted) is a set of \texttt{tableIDs}, which can later be used
in requesting raw data.  The result for \texttt{queryDataScheme} is a
set of tuples \texttt{(columnID, type)} representing the column name
and type.

\subsubsection{Querying data.}
Clients can request data by invoking:
{\small
\begin{verbatim}
{{data record}, {matching policies}, fail} 
             <- queryData(requested resources,
                          joining condition)
\end{verbatim}
}
where
{\small
\begin{verbatim}
requested resources 
           = {<credentials, dataId, {columns},
              {actions}, {constraints}>}
\end{verbatim}
}
represents the resources requested from different datasets.
\texttt{joining condition} specifies how the results from those
datasets are joined. These results are returned separately if
\texttt{joining condition} is \texttt{null}. \texttt{constraints}
contains conditions that are applied to the returned data. For
example, $column_i > \theta$ where $column_i \in
\{\mbox{\texttt{columns}\}}$ indicates that the request is only for
data whose $column_i$ values are greater than $\theta$. The protocol
proceeds as follows:
\begin{enumerate}
	\item For every requested resource, the proxy creates a
	well-formed XACML request using \texttt{dataId},
	\texttt{columns} as \emph{Resources} and \texttt{actions} as
	\emph{Actions} attributes. The request is then forwarded to
	the server specified by \texttt{dataId}.
	
      \item The XACML* instance returns access control decision, the
        accompanied data (if decision permitted), and IDs of the
        matching policies.

      \item The proxy, on receipt of non-empty data, applies
        conditions specified in \texttt{contraints}. Depending on the
        value of \texttt{joining column}, it performs data joining
        (discussed next) before sending the final response to the
        client.
\end{enumerate}

\subsection{Data Joining.}
The \texttt{joining condition} parameter used in \texttt{queryData}
specifies how the results are joined before returning to the client.
In particular:
\[
\mbox{\texttt{joining condition}} \in \{\mbox{\texttt{null}}, \{c_1, c_2, .., c_k\}\}
\]
where $k$ is the number of requested resources and $c_i$ $(1 \leq
i \leq k)$ are the joining columns of the returned data. When
\texttt{joining column = null}, the proxy forwards what it receives
from the server directly back to the client. Otherwise, it waits until
getting data from all requested servers, then constructs a client
response by joining the results using normal database join operations.

\subsection{Conflict Resolution.}
It is possible for clients to receive empty data for their requests,
especially when the requests involve more than one datasets. This
arises because different policies associated with different datasets
are enforced. We refer to this as \emph{policy conflict}, which
happens in one of the two cases:
\begin{enumerate}
\item There is at least one policy that denies the client's access.
\item All policies permit access, but the joined data still results in
  an empty set. For example, one policy allows access to data where
  $column_i > \theta$ whereas another policy allows access to data
  where $column_i \leq \theta$. Another example is when two policies
  specify different sliding windows, as a consequence the joining columns
  do not have values in common.
\end{enumerate}

We provide a simple mechanism for dealing with policy conflict.
Responses from \texttt{queryData} includes IDs of the matching
policies. When conflict occurs, the client is aware of the cause and
is able to contact the dataset owner to resolve the conflict. We assume
that such resolution is done out-of-band and is not within the scope
of the framework.

\subsection{Caching.}
The proxy maintains a cache of data received from the servers. Since
operations in the cloud server are slow, especially when involving
database access, caching can improve the response time.  It is also
reasonable to expect a cache-friendly request pattern from clients,
as popular data are frequently requested.

We consider a simple design, in which data cache is the map
\texttt{<request>:<data>} where \texttt{request} is the XACML request
with the corresponding \texttt{data}.
\begin{itemize}
	\item \emph{Cache replacement}: when full, an old entry is
	evicted in a random fashion.

      \item \emph{Cache coherence}: stale entries can lead to security
        violation. For instance, a new policy update denies a client
        access to a dataset, but the cache contains data of previous
        access which will be served by the proxy at the client's next
        request. We address this problem by simply purging entire
        cache every time a policy is loaded or removed.
\end{itemize}

\section{Prototype and Evaluation}
\label{sec:evaluation}
\subsection{Prototype}
We have implemented a prototype of eXACML, which consists of over
$3400$ lines of Java code. Database accesses are provided by JDBC API,
while communications between clients, proxy and servers are done
through Socket interface. For XACML*, we extended Sun's XACML
implementation~\cite{sunxacml} --- an open source, Java project that
supports XACML 2.0 standard. We instrumented its PEP module to handle
more obligations (Section 3). The prototype supports all the features
discussed in the previous section: a client is able to load, remove,
query data and policies.

\begin{figure}
\centering
\subfloat[Data view]{\includegraphics[scale=0.4]{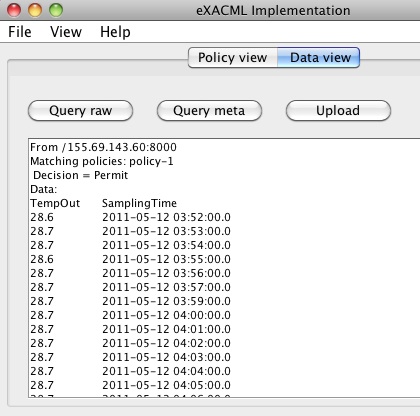}}
\ \ \
\subfloat[Query form]{\includegraphics[scale=0.3]{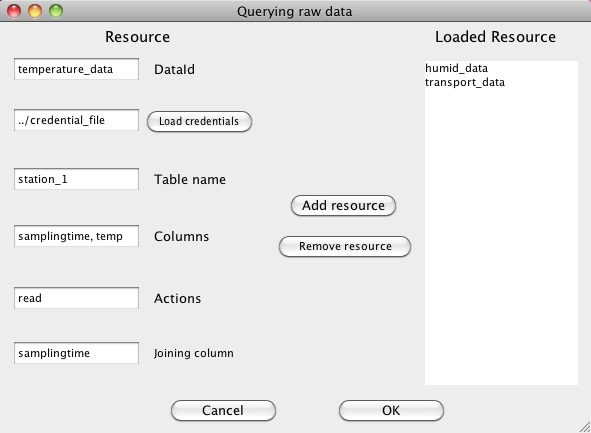}}
\caption{User interface for querying data}
\label{fig:gui_querying}
\end{figure}
\begin{figure}
\centering
\subfloat[Policy view]{\includegraphics[scale=0.4]{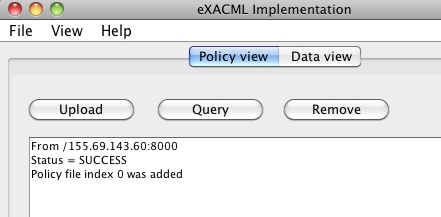}}
\ \ \
\subfloat[Policy upload]{\includegraphics[scale=0.45]{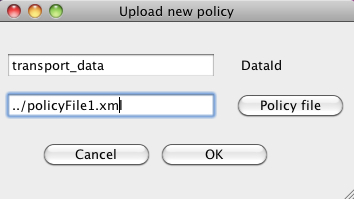}}
\caption{User interface for managing access control policies.}
\label{fig:gui_management}
\end{figure}

Our prototype provides an easy-to-use graphical interface for querying
and managing data. A query form (Fig.~\ref{fig:gui_querying}b) takes
in user credentials and requests.  A response from the server includes
the data server information, matching policies and the data (if
applicable), which are displayed in the data view window
(Fig.~\ref{fig:gui_querying}a).  Policies are updated and queried
using similar GUI, as shown in Fig.~\ref{fig:gui_management}.

\subsection{Evaluation}
We evaluated our prototype's performance, and its ability to
support dynamic, fine-grained access control policies. The system
performance is measured by the time taken to fulfill user requests. We
compare our prototype's performance against that of a system that
executes the requests directly, i.e. without the access control layer.
We refer to the later as \emph{direct-query system}.

\subsubsection{Methodologies. }
\paragraph{\textbf{Setup.}}
We emulate a cloud-like environment running our prototype, as shown
in Fig.~\ref{fig:framework}. More specifically, we make use of four
machines, two running servers, on running the proxy and the other
represents a client. The machines belong to the PDCC
cluster\footnote{\url{http://pdcc.ntu.edu.sg/content/128-cores-linux-cluster-pdccsce}},
each has one Xeon processor 3.0Ghz, running OCS5.1 (2.6.18-53El5smp)
operating system with 4GB of RAM. The machines are connected via
InfiniBand 20Gbps.

The servers maintain two databases: a weather database and a traffic
database. The former contains four tables with real data taken from
four different weather stations collected in a 5-day duration and with
one-minute sampling interval. We synthesize the traffic database with
two tables containing records of traffic volume and vehicle speed that
match with the weather datasets.

\paragraph{\textbf{Workloads.}}
We generate synthetic workloads that include large numbers of policies
and requests. Since our prototype is compared against a direct-query
system, the workloads also
contain a large number of direct database queries, each corresponds to
a request in our prototype. A \emph{direct query} is forwarded
to the server, which executes and returns the same data as when
executing the corresponding request in our system. The parameters
used in generating workloads are shown in Table.~\ref{tab:variables}.
The workloads and source code for generating them can be found at
\url{http://sands.sce.ntu.edu.sg/trac/exacml/}

First, we use $nDirectQueries$ and $directQueryDist$ to create a set
$DQuery$ of direct queries of five different types: selection,
approximation, aggregation, sliding window and data joining. The first
three types are ordinary database SELECT query, which is forwarded by
the server directly to the database engine. Sliding window queries are
first converted into multiple SELECT queries, one for every window,
which are then sent to the database engine. Data joining queries
contain two sub-queries (of the other four types) chosen at random and
for different data servers. Each data server processes and returns the
result independently.  Next, $nPolicies$ unique XACML policies are
generated, each with different \texttt{exacml.subject:role-id}. Every
policy corresponds to a direct query whose type is either selection,
approximation, aggregation or sliding window. Therefore, the set of policy
obligations and $DQuery$ represent the same set of SELECT queries to
be executed by the database engines.

\begin{table*}
\centering
\scriptsize
\begin{tabular}{|l|l|p{4.5cm}|}
  \hline
  Variable & Value & Description\\ \hline

  $nDirectQueries$ & 1000 & number of direct queries \\ \hline

  $directQueryDist$ & 248:248:248:156:100 & distribution of direct
  queries (selection:approximation:aggregation:sliding window:joining
  request) \\ \hline

  $nPolicies$ & 900 & number of unique policies \\ \hline

  $nRequests$ & 1500 & number of matching requests \\ \hline


  $\alpha$ & 0.223 & skew parameter for Zipf distribution\\ \hline

  $maxRank$ & 300 & maximum rank of unique requests from which Zipf
  distribution is generated \\ \hline
\end{tabular}
\caption{Summary of parameters used in setting up experiments}
\label{tab:variables}
\end{table*}

Next, we generate a set of requests. For every policy, we construct
one matching and one non-matching request. The matching request
contains credentials, resources and actions as specified in the
policy. For the non-matching request, we use a different\\
\texttt{exacml:rdbms-database-id} from the weather and traffic
database names. For each data joining direct query, we create
corresponding (matching and non-matching) requests made up of two
sub-requests. Each sub-requests from the matching request corresponds
to a sub-query in the data joining direct query. In summary, a
matching request executed in our prototype returns the same data as
the corresponding query evaluated in the direct-query system.

Finally, we create a workload of $nRequests$ requests following Zipf
distribution with skew parameter $\alpha$. This workload models a
realistic use of the prototype, in which a small number of popular
data are requested frequently. Such request pattern is found in many
other systems, such as P2P file-sharing and web caching
\cite{adamic02,klemm04}. We select $maxRank$ unique queries from
$DQueries$ at random, then assign them with random ranks. A sequence
of queries is generated from the selected set with Zipf distribution,
using $\alpha=0.223$ (as in \cite{klemm04}). For every direct query,
this workload also contains the corresponding policy, matching and
non-matching request.

\subsubsection{Metrics.}
In the following experiments, we investigate our prototype's
effectiveness in granting data access to authorized requests and
denying unauthorized ones. We also measure its performance in terms of
the time taken to fulfill authorized data requests. This is compared
against the direct-query system, i.e. one without eXACML. We also
provide quantitative analysis of the proxy, especially its caching and
data joining features.

\subsubsection{Experiments and Results.}
We first load $nPolicies$ unique policies onto the data server. The
measured time is reasonably small, with mean of $0.034s$ and standard
deviation of $0.016$ per loading operation.

We then run two sets of experiments:
\begin{enumerate}
\item The workload consisting of $nDirectQueries$ unique queries and
  the corresponding unique requests. We enable the data joining option
  at the proxy in the first run, and disable it in the second. To
  disable cache, we simply change the proxy configuration file. To run
  without the joining option, we re-generate the workload without data
  joining queries and requests. We measure the time taken to fulfill
  direct queries and data requests.
\item The workload contains $nRequests$ queries and the corresponding
  requests, which follow the Zipf distribution.
\end{enumerate}

\begin{figure}
\includegraphics[angle=-90, scale=0.35]{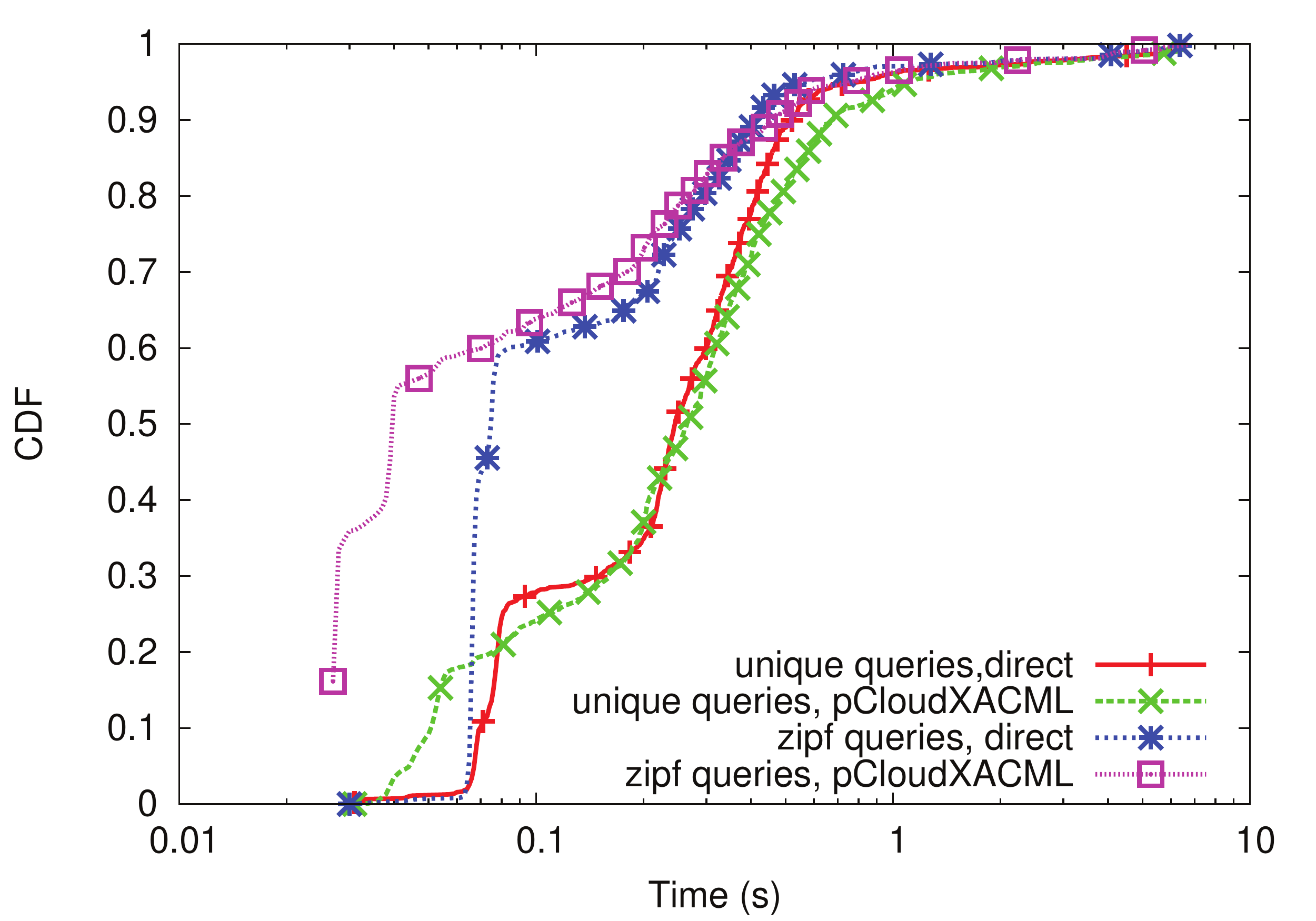}
\caption{Overall performance, with vs without exacmlXACML. Caching and joining options are on}
\label{fig:overall}
\end{figure}
\begin{figure}
\includegraphics[angle=-90, scale=0.35]{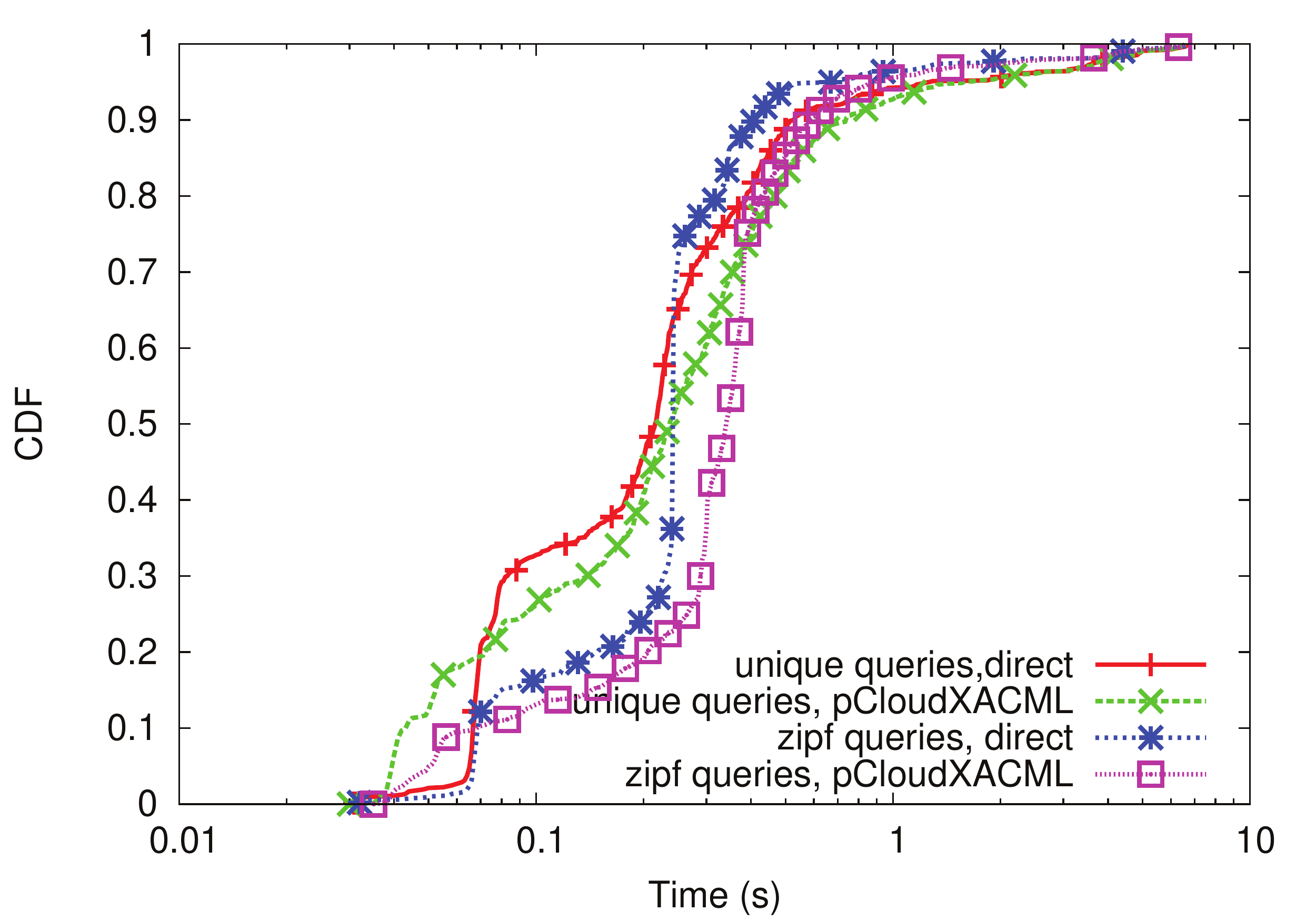}
\caption{Overall performance with exacmlXACML when the joining and caching options are disabled}
\label{fig:overallOptionsOff}
\end{figure}
In both experiments, non-matching requests are denied access.
Fig.~\ref{fig:overall} and Fig.~\ref{fig:overallOptionsOff} compare
the performance of our prototype against direct-query system, using
measurements of matching requests. In both figures, there is a number
of requests taking over $5s$ to finish. They are sliding window
requests, which translates into a large number of SELECT queries to be
executed by the database engines.  That the server needs to wait and
aggregate the results into a single client message, and that JDBC
implementation incurs non-significant overhead for executing a SELECT
query both contribute to the noticeable delay.

Fig.~\ref{fig:overall} illustrates eXACML's overhead when both caching
and data joining options at the proxy are enabled. For unique queries
and requests, there is no overhead from the $99^{th}$ percentile.
$80\%$ of the requests incurs less than $10\%$ overhead. The largest
overhead is less than $0.4s$ and is observed from between $87\%$ to
$90\%$ percentile. An interesting pattern in which eXACML outperforms
the direct-query system can be seen at lower percentiles. Besides
network and computational variations, this can be attributed to the
data joining feature at the proxy (discussed later). For requests and
queries following Zipf distribution, eXACML performs better most of
the time (up until the $89^{th}$ percentile).  This is thanks to the
caching mechanism at the proxy, whose benefit will be analyzed in more
detail later.

Fig.~\ref{fig:overallOptionsOff} shows how the overhead changes when
the proxy performs neither caching nor data joining.  The overhead is
more discernible: for unique requests, the overhead starts from
$20^{th}$ percentile, as compared to $45^{th}$ percentile in
Fig.~\ref{fig:overall}. Similarly, for queries following Zipf
distribution, the overhead is seen from $10^{th}$ percentile, as
compared to $89^{th}$ percentile in Fig.~\ref{fig:overall}. This
implies that caching and data joining at the proxy are most effective
when the query distribution is heavy-tailed.

\begin{figure}
\includegraphics[angle=-90, scale=0.35]{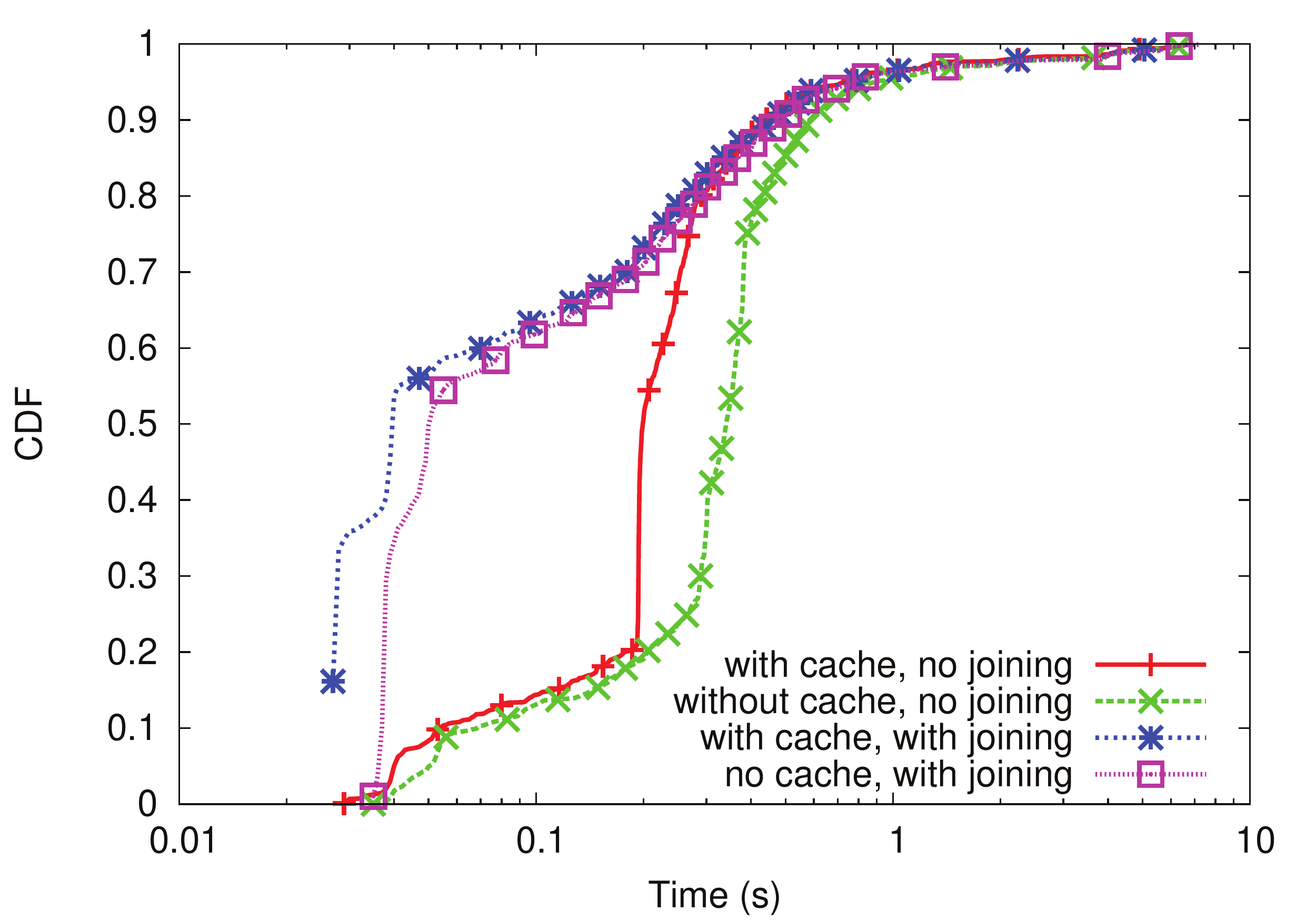}
\caption{Benefit of caching on performance. Queries follow Zipf distribution}
\label{fig:caching}
\end{figure}

We proceed to analyze benefits of caching at the proxy. Request times
for Zipf-distribution queries with and without cache are extracted
from the experiments and plotted in Fig.~\ref{fig:caching}. We show
the results with and without data joining queries. In both cases,
caching results in better performance. By itself, i.e. without the
joining data feature, caching leads to $50\%$ improvement for more
than $80\%$ of the requests. For the workload including data joining
queries, a similar pattern can be seen, although the improvement is
not as noticeable.

\begin{figure}
\includegraphics[angle=-90, scale=0.35]{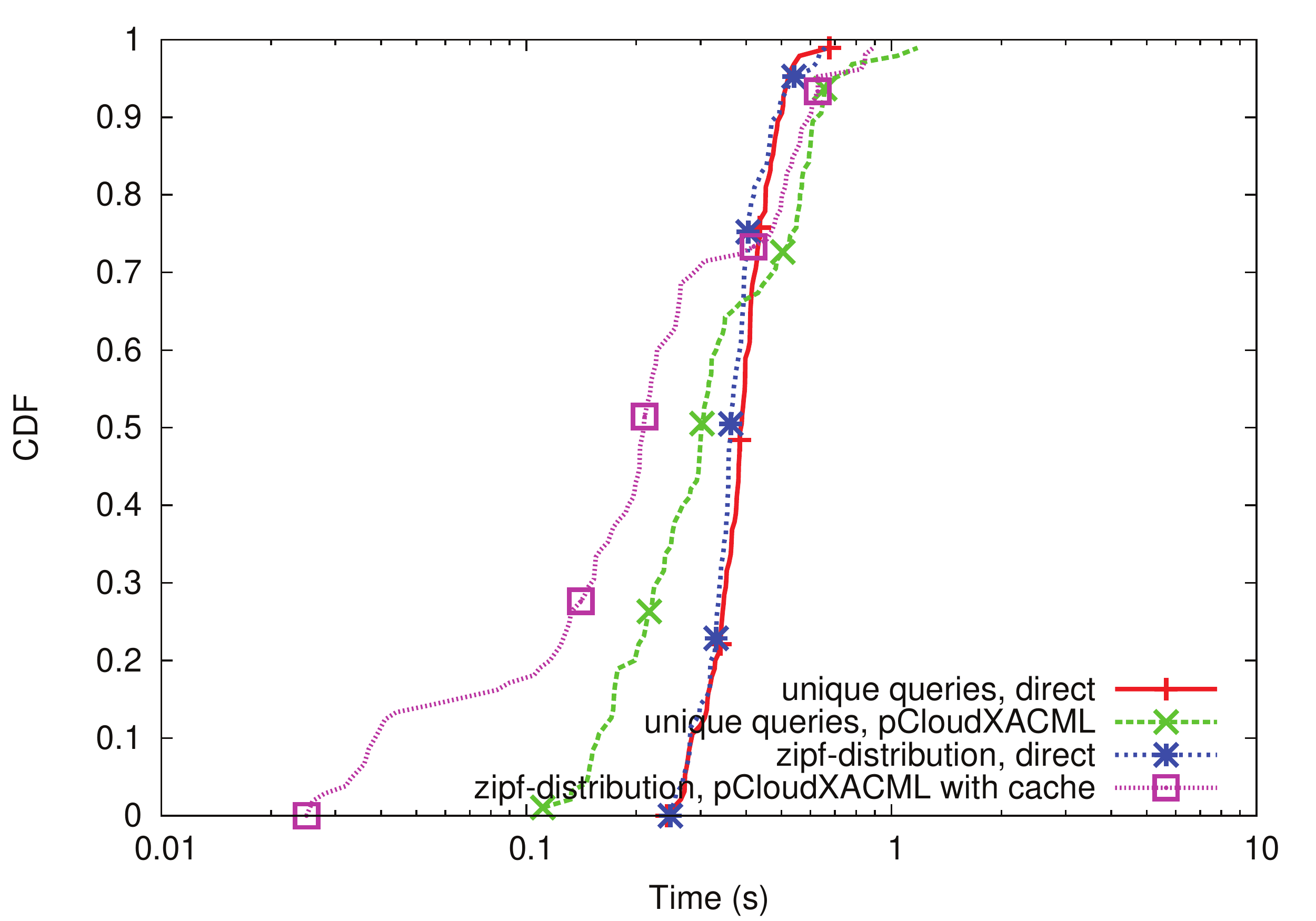}
\caption{Benefit of proxy performing data joins. All queries require data joining}
\label{fig:joining}
\end{figure}

Finally, we analyze the benefit of the data joining feature at the
proxy. We run the same experiments as before, but with workloads
consisting of only data joining queries and requests. The results shown
in Fig.~\ref{fig:joining} are for both unique and Zipf-distribution
requests. It can be seen that eXACML outperforms the direct-query
system up until $65^{th}$ percentile for unique queries and $70^{th}$
percentile for Zipf-distribution queries.  This is because for most
requests, eXACML helps reducing the data size substantially (by
joining the results from two servers) before transferring it back to
the client. In contrast, without eXACML, the client has to wait for
all data to come back individually before performing joining by
itself. Notice that some requests in eXACML still experience longer
delay (after $70^{th}$ percentile), because extra communication
between client and proxy (as opposed to the direct communication
between client and server) and computation overhead at the proxy are not fully
discounted.

\section{Related work}
\label{sec:relatedwork}
There exists cloud-based systems that enable data sharing from
multiple sources. SenseWeb~\cite{senseweb},
SensorBase~\cite{sensorbase} are examples of cloud services that let
users upload and share their sensor data. They support coarse-grained
access control model in which an user either makes its dataset public,
shares it with a list of collaborators or keeps it private. Similarly,
Google's Fusion Table~\cite{fusionTable} allows user to upload generic
data and to perform simple analysis such as data visualization on the
cloud. Recently, companies such as Okta~\cite{okta} have started
implementing cloud-brokerage models that provide centralized service
for management of enterprises' resources, including access control.
However, these access control model is also coarse-grained, which
means it cannot deal with the access scenarios we consider in this
paper. In addition, data owners in these systems upload their datasets
onto a centralized cloud, whereas our work does not make such
assumption (we consider multi-cloud environment in which different
data owner uses its own cloud provider).

There are also numerous works focusing on access control and data
privacy on the cloud. Airavat~\cite{roy10}, for example, assumes the
cloud is trusted in enforcing access control. It uses a simple
mandatory access control system available in SELinux~\cite{selinux},
and provides a trusted environment for executing
MapReduce~\cite{dean04} jobs while guaranteeing differential
privacy~\cite{dwork06}. Our work makes the same assumption about
clouds' trustworthiness, but aims at improving the access control
aspect of the system, which is complementary to Airavat. Other
works~\cite{yu10,kallahalla03,popa11} assume the cloud is untrusted
and employ cryptographic approach for access control.  In~\cite{yu10},
data is encrypted with attribute-based
encryption~\cite{goyal06,bethencour07} by a proxy using a proxy
re-encryption technique. Embedded in the ciphertext are conditions
that must be met when decrypting. Plustus and
CloudProof~\cite{kallahalla03,popa11} use broadcast
encryption~\cite{naor01} to protect the data, while key
management~\cite{kallahalla03} is done using key rolling and lazy
revocation techniques. These cryptographic approaches provide strong
guarantees for data security, but they cannot express fine-grained
access control policies as described in our work. Thus the focus in these works is also complimentary to ours. In addition, key
management and revocation protocols are complex and incur much
overhead in such an untrusted environment.

Multiple policies matching in XACML is usually resolved by the
top-level policy combining algorithms. XACML supports only a limited
number of combining algorithms. Ninghui et al.~\cite{ninghui09} and
Rao et al.~\cite{rao09} propose a formal language for expressing more
fine-grained policy composition. The language can deal with evaluation
errors and combining of obligations. Mazolleni et
al.~\cite{mazzoleni06} propose a method for combining policies based
on their similarity and users' preferences.

Time-series data --- similar to those considered in our paper ---
could arrive at the system in continuous streams, for which relational
databases such as MySQL and Postgresql are not ideal.
Aurora~\cite{abadi03} is a popular data stream management system that
addresses limitations of relational databases when it comes to stream
data. Carminati et al.~\cite{carminati07,carminati07a} are among the
first to propose a model and implementation of access control for data
streams based on Aurora. The model supports four access scenarios:
column-based, value-based, general window and sliding window. Our
framework supports all of these scenarios for on-demand queries over
archival databases. The extension to eXACML that deals with continuous
queries over stream databases is left for future work.

\section{Future work}
\label{sec:future work}
We have implemented a simple prototype and carried out preliminary
evaluation of our framework. The next step would be to improve the
prototype and perform more comprehensive evaluations. More specifically,
the cloud-like environment set up in the experiment contains only two
data servers. In addition, only one dataset comes from real monitoring
stations, and the workloads are synthetic. Therefore, we plan to
acquire more realistic datasets and workloads, and to evaluate the prototype
with larger numbers of data servers. We also plan to export our
prototype into real cloud environments such as Amazon EC2 and
Microsoft's Azure~\cite{ec2,azure}, and benchmark it with real data
mining applications accessing real datasets.

We assumed that each dataset is guarded by an independent XACML*
instance. We have acknowledged the trade-offs in having multiple
datasets sharing one XACML* instance, especially when datasets reside
in the same physical machine. Another trade-off is the number of
proxy servers. It would be interesting to investigate these trade-offs
further by extending the framework with XACML* sharing and distributed
proxies.

As shown in Table~\ref{tab:scope}, eXACML only deals with archival
databases and queries. The immediate extension will be to support
stream databases and continuous queries. Relational databases are not
the best tool for handling stream data, for which other models have
been proposed~\cite{abadi03}.  We will examine the design and compare
performance of the extended eXACML to that of the existing works on
access control for stream data~\cite{carminati07,carminati07a}.

Regarding data sharing, access control only addresses the problem of
authorization. We have so far made an assumption that authentication
is implicit, that is, clients are given static credentials and the
servers always accept the given credentials. We plan to incorporate an
authentication model into our framework. It is an interesting
challenge in decentralized settings, of which our multi-cloud scenario
is an example, since authentication may depend not only on static
credentials but also on previous interactions between parties and the
states of the entire system. Authentication is also an important when
the cloud provider has to log and notify data owners of access to
their data (for billing purposes, for example). We plan to use other
access control languages such as DynPal~\cite{dynpal} or
SecPal~\cite{secpal}, because they are more suitable for handling
dynamic authentication than XACML.

Finally, we have always assumed the cloud is trusted in enforcing
access control policies and not to violate user's data security and
privacy. However, users with sensitive data or data that have been
expensive to collect will demand highest level of security. As a
consequence, they cannot assume the cloud is trusted in handling their
data. Existing works have taken the cryptographic approach that
encrypt data and attempts to outsource the key management to the
cloud.  Nevertheless, the range of access control policies supported
by the existing systems has been limited. For future work, we aim to
find practical cryptographic protocols that can handle more
fine-grained access control scenarios. Since eXACML contains two
components belonging to third parties: the proxy server and the cloud
servers, we will investigate relaxing the trust assumption for these
components one by one.

\section{Conclusion}
\label{sec:conclusion}
In this paper, we have proposed a framework (eXACML) that allows users to share
their data on the cloud in a secure, flexible, easy-to-use and
scalable manner. We considered a trusted cloud environment, in which
data are maintained in relational databases. The cloud environment
makes it easy for data owners to share and benefit from mining the
aggregated data.  The main challenge is how to let users control
access to their data in most flexible ways. We achieved security
and flexibility by extending the XACML framework, allowing users to
specify fine-grained access control policies. Our framework contains
a proxy server residing in between clients and the cloud servers.
It processes requests from the clients, joins and caches responses from
the servers before sending back to the client. We have implemented a
prototype and carried out preliminary experiments to
evaluate its performance. The results suggested that the framework is
scalable, as the overhead incurred is small, thanks to the caching and
data joining features at the proxy. In addition, the prototype
provides a graphical user interface that lets users share and manage
their data in an easy-to-use manner.

We believe that in order to take full advantage of cloud computing,
having a framework such as ours is very important. Our paper has taken
the first steps towards realizing a practical and usable
sharing-friendly cloud environment. We have also identified many avenues
for future work, such as improving scalability with more proxies,
adding support for stream data and other policy languages, and relaxing
assumptions on the trustworthiness of the cloud.

\paragraph{Acknoledgments.}
  This work has been supported by A*Star TSRP grant number 1021580038
  for `pCloud: Privacy in data value chains using peer-to-peer
  primitives' project. The authors will like to thank Dr. Lim Hock
  Beng for providing access to the weather data sets used in some of
  the experiments.

\bibliographystyle{IEEEtranS}       
\bibliography{pCloud}   

%
%

\end{document}